# The Gemini Planet Imager: Looking back over five years and forward to the future

Bruce Macintosh[a], Jeffrey K. Chilcote[a,b], Vanessa P. Bailey[c], Rob De Rosa[d], Eric Nielsen[a], Andrew Norton[a], Lisa Poyneer[e], Jason Wang[d], J.B. Ruffio[a], J.R. Graham[d], Christian Marois[f], Dmitry Savransky[g], and Jean-Pierre Veran[f]

[a]Kavli Institute for Particle Astrophysics and Cosmology, Stanford University, Stanford, CA 94305, USA;
[b]Department of Physics, University of Notre Dame, 225 Nieuwland Science Hall, Notre Dame, IN, 46556, USA;
[c]Jet Propulsion Laboratory, California Institute of Technology, 4800 Oak Grove Dr., Pasadena CA 91109, USA;
[d]Department of Astronomy, UC Berkeley, Berkeley CA, 94720, USA;
[e]Lawrence Livermore National Laboratory, 7000 East Ave., Livermore CA 94551, USA;
[f]National Research Council of Canada Herzberg, 5071 West Saanich Road, Victoria, BC V9E 2E7, Canada;
[g]Sibley School of Mechanical and Aerospace Engineering, Cornell University, Ithaca, NY 14853, USA

## ABSTRACT

The Gemini Planet Imager (GPI), a coronagraphic adaptive optics instrument designed for spectroscopy of extrasolar planets, had first light in 2013[13]. After five years, GPI has observed more than 500 stars, producing an extensive library of science images and associated telemetry that can be analyzed to determine performance predictors. We will present a summary of on-sky performance and lessons learned. The two most significant factors determining bright star contrast performance are atmospheric coherence time and the presence of dome seeing. With a possible move to Gemini North, we are planning potential upgrades including a pyramid-sensor based AO system with predictive control; we will summarize upgrade options and the science they would enable.

## 1. INTRODUCTION

The Gemini Planet Imager (GPI) is an integrated instrument designed for the detection and characterization of young self-luminous giant planets and circumstellar debris disks. It is used as both a facility instrument, with 700 hours of guest observer programs accepted to date, and for a large-scale exoplanet search campaign, the Gemini Planet Imager Exoplanet Survey (GPIES), which was allocated 895 hours.

It combines an adaptive optics (AO) system, an apodized-pupil Lyot coronagraph (APLC), a near-infrared interferometric wavefront sensor (CAL) and a near-infrared integral field spectrograph (IFS). An opto-mechanical subsystem (OMSS) mounts the individual subsystems in a truss structure and inside a clean enclosure. High-level software (TLC) coordinates the subsystems during operation. Finally, a sophisticated data pipeline processes and calibrates raw 2D IFS images into 3D data cubes. This publically-available data pipeline has been supplemented with a complete data infrastructure for the GPI Exoplanet Survey (GPIES) campaign, which incorporates automatic point spread function (PSF )subtraction and archiving of both science images and telemetry/metadata. GPI achieved first light in November 2013 and began science operations one year later. In this paper we will review key features of the instrument and lessons learned from them, discuss performance and the factors that determine it, and briefly discuss potential upgrades.

## 2. SUBSYSTEMS OF THE GPI INSTURMENT

### 2.1 Adaptive optics

The GPI adaptive optics system uses a Shack-Hartmann wavefront sensor that has 43 subapertures across the 7.8-m effective diameter of the Gemini South telescope. Each subaperture is imaged onto a 2x2 pixel quadcell on a 128x128 pixel active area of a Lincoln Labs CCD. Wavefronts are reconstructed using a Fourier-transform based algorithm[1] on a commercial multicore Xeon-based computer. Phase corrections are sent to a 48x48 actuator subsection of a 4096-actuator Boston Micromachines MEMS deformable mirror ("tweeter") and a 9-actuator-diameter conventional lower-order piezeo-electric MEMS mirror ("woofer"), as well as a tip/tilt stage. The AO system operates at 500 or 1000 Hz, with a total delay from start of CCD readout of 1.6 ms. With the conventional wavefront sensor CCD, GPI has a limiting AO magnitude of I=10. The MEMS deformable mirror has 5 defective actuators within the pupil, which are blocked in the coronagraph Lyot plane.

Tip/tilt control is provided by the woofer. High temporal frequencies are sent to the actuators controlling the woofer surface, merged with other low-order wavefront errors. Lower temporal frequencies are sent to a piezoelectric stage moving the entire woofer assembly. The total range of this stage is relatively low (2 arcseconds), which can become a performance issue during times of high windshake. The tip/tilt system incorporates an LQG-based vibration-rejection mode which can be set to three independent frequencies; this was used to reject 60 Hz and harmonic vibrations from GPI's cryocoolers before their anti-vibration upgrade (Section 2.4) as well as 37 Hz vibrations from the telescope secondary, but is currently rarely needed.

Performance of the adaptive optics system has been analyzed in previous papers[2]. The most significant predictor of AO performance, as measured by the wavefront error seen directly by the wavefront sensor, is (unsurprisingly) the atmospheric coherence time $\tau_0$. Measurements of free-atmosphere $\tau_0$ are available from a MASS sensor that reports a median value of 1 ms. While the absolute calibration of the sensor could be in error by ~50%, it is clear that typical $\tau_0$ are substantially faster than the model atmosphere provided for the GPI design (~5 ms); combined with the relatively long total delay, GPI servo-lag errors are roughly a factor of two worse than expected, which in results in a brighter PSF halo, especially in the 0.2-0.6" range. Dome seeing is also a significant contributor to residual wavefront error[3].

A key feature of GPI are high-quality optics; individual off-axis aspheres in the AO optical path were polished to ~1 nm RMS phase error at mid-frequencies. This minimizes both Fresnel chromaticity effects and non-common-path errors.

### 2.2 Coronagraph

GPI's primary coronagraph is an apodized-pupil Lyot coronagraph[4]. Selectable pupil-plane apodizers are available for each wavelength band, combined with matched focal-plane occultors and Lyot stops. Since GPI operates in a single *YJH* (or partial *K*) band at any given time, the APLC can be well-matched to the observing wavelength, producing achromatic PSFs with relatively high throughput. Occulting masks have a radius of 2.8 λ/D, setting a hard inner working angle (IWA) limit of 0.12" at *H* band, but the APLC design results in a bright Airy-like ring surrounding the mask so the practical IWA is ~0.15". APLCs designed with modern numerical optimization could potentially mitigate this issue. The apodizers patterns also include a coarse grid that produces four "satellite spots"[5][6] in the image, intended to track the intensity and location of the central star. In operation these spots are somewhat distorted by unknown optical effects (possibly the interaction between the grid and mid-spatial-frequency aberrations), but can still typically be centroided to ~1 miliarcsecond and photometrically measured to ~5%.

The coronagraph apodizer wheel also includes a newly-added ND3 filter for bright star observations and a non-redundant aperture mask[7].

## 2.3 Infrared wavefront sensor

GPI's infrared wavefront sensor module, referred to as the CAL subsystem, was intended to measure the near-IR wavefront at the coronagraph PFM to minimize non-common-path and chromatic errors. It includes both a low-order Shack-Hartmann sensor (LOWFS) using light from the focal plane mask and a high-order Mach-Zehnder interferometer (HOWFS) using the light from the off-axis science beam. Neither is fully operational. In practice, the residual low-order aberrations are larger than the calibration accuracy of the LOWFS, since no provision exists to feed it with a beam matching the GPI pupil. Instead, it is used to align stars to the focal plane mask and to measure tip/tilt to retain that alignment. The HOWFS interferometer is sensitive to vibration and does not produce detectable fringes under telescope conditions. Fortunately, the intrinsic mid-frequency NCP errors of GPI are relatively small (< 5nm, dominated by WFS lenslet and CCD effects) so even without the HOWFS PSFs are stable during good seeing. The CAL box also houses the focal plane masks and a collimating optic to produce the beam feeding the spectrograph.

## 2.4 Integral field spectrograph

GPI's science instrument is a low spectral resolution (R~40 at *H* band) lenslet-based integral field spectrograph[8]. The imaging field of view is 2.8x2.8", and images are taken in a single standard IR band at one time. Spectral resolution increases with longer wavelengths so the *K* band at R~70 is split into two sub-bands. The H2RG detector is used in up-the-ramp sampling mode with ~5 electrons readnoise in a 60-second-equivalent exposure. The IFS also includes a polarimetric mode in which the spectral prism is replaced by a Wollaston prism, giving two broad-band images in orthogonal polarizations. Combined with a rotating waveplate this leads to a complete measurement of the linear Stokes components of each image pixel, and high sensitivity through differential polarimetry to circumstellar disks. The current design does not allow spectropolarimetry.

The IFS is cooled by two Sunpower Stirling-cycle cryocoolers. These produced significant 60 Hz vibration – up to 10-20 mas of image motion both internal to GPI and through vibration of the telescope. The LQG-based tip/tilt loop helped to reject these as did adjustments to the phase of the cryocoolers, but ultimately the cryocoolers were upgraded with active vibration damping and residual tip/tilt is now below 5 mas.

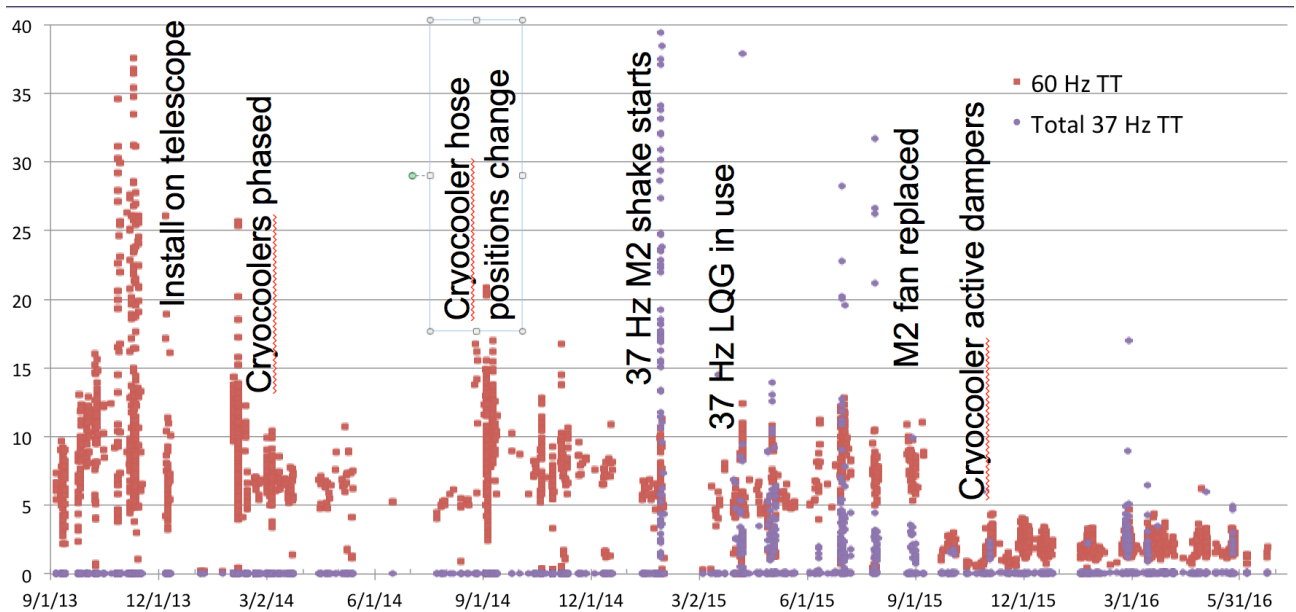

**Figure 1: GPI measured closed-loop image motion at 60 and 37 Hz through time. 60 Hz vibration was induced by the GPI cryocoolers while 37 Hz vibration was induced by a bad fan exciting the Gemini secondary mirror M2. Vibrations were attenuated by the GPI LQG controller and ultimately prevented by repair of M2 and addition of active dampers on the cryocoolers.**

## 2.5 TLC software

The Top Level Computer (TLC) controls the mechanisms on the GPI optical bench and commands the individual subsystems. It includes an instrument sequencer that expands observatory commands (such as "configure for *H* coronagraph mode" or "close all loops") into a sequence of subsystem commands. In principal this should allow "one-button" operation from the observatory software[9], but in practice the process is somewhat more complicated due to the limitations of the Gemini instrument sequencer and the need for human checks.

## 2.6 Data pipeline

Raw IFS data are two-dimensional IR array images containing a 200x200 grid of individual microspectra corresponding to each IFS lenslet. The GPI data pipeline[10] converts these into three-dimensional data cubes calibrated in wavelength to an accuracy of 2 nm using an argon arc lamp. It also carries out standard data quality checks, identifying the "satellite spots", evaluating the image contrast at 0.2, 0.4, and 0.8" separation, marking saturated pixels, etc. The data pipeline runs in IDL, either through a local IDL license or as (free) precompiled IDL modules.

All GPI data passes through the Gemini Data Handling System and is available through the Gemini archives. Image headers are populated with observatory and GPI telemetry such as temperature or residual AO WFE. The GPIES campaign maintains is own independent archive/database of campaign observations, associating header keywords and external data (such as additional weather information) with each image. The database also stores telemetry from the AO system and AO analysis products. A typical GPIES sequence on a target star consists of 40 images of 60 seconds exposure each, taken in angular differential imaging (ADI) mode. GPIES data is processed through an extended architecture, built around the data pipeline, referred to as the Data Cruncher. This combines the IFS processing with full PSF subtraction in both angular and spectral differential using the KLIP algorithm. Partially-processed PSF subtracted images with several algorithms are available in real time, and KLIP processing of a complete set is typically finished 10-20 minutes after the end of an observing sequence. Final images are automatically searched for planets using a matched-filter algorithm, and estimated contrasts are entered into the database.

This architecture is one of the most successful aspects of the GPI / GPIES program. It allows observers to identify candidate companions and examine their spectra in near-real time and prioritize follow-up observations. (For example, anomalies in the first GPI spectrum of the putative exoplanet HD131399b[12] were identified within an hour of observation.) It is straightforward to assess the total sensitivity and status of the campaign, to perform data checks on new observations, or to correlate GPI performance with database parameters (see section 3.2.)

# 3. PERFORMANCE

## 3.1 Scientific highlights

Since 2014 there have been 46 papers based wholly or in part on GPI data, roughly split between GPIES and guest observer projects. There are many significant exoplanet results including the discovery of 51 Eridani b[13], detailed spectral and orbital characterization of other planetary companions, and discovery and characterization of brown dwarfs. GPI's polarimetry mode has been applied to extensive observations of circumstellar disks.

The GPIES campaign has executed 715 hours out of 895 allocated. The campaign was originally intended to complete in 3 years, but Chilean weather has been exceptionally poor; the campaign will stop searching at the end of 2018 with (likely) less than the full 600 stars surveyed. Overall, exoplanet discovery rates with direct imaging by GPIES have been significantly lower than early forecasts[15], similar to the results of surveys with SPHERE and earlier facilities[16]. Statistical interpretation of the GPIES survey is in progress, but the discovery rates likely indicate that the giant planet distribution, especially for solar-type stars, do not continue to rise with increasing semi-major axes beyond ~5 AU; though average occurrence rates are consistent with those measured from 0.3-3 AU. It is also possible that a significant number of planets exist that are either formed with low entropy "cold start" initial conditions[17] or have masses below 2 MJ, where current facilities have little sensitivity.

## 3.2 Performance and correlations

The figure of merit for high-contrast imaging is "contrast", defined generally as the flux ratio at which a companion object could be detected as a function of angular separation from a target star. Typically, this is set as multiple of the local standard deviation at a given radius after appropriate processing such as PSF subtraction, collapse of spectral channels, matched filtering[17], etc. Figure 2 shows the median contrast from GPIES campaign sequences computed using the forward-model matched-filter approach.

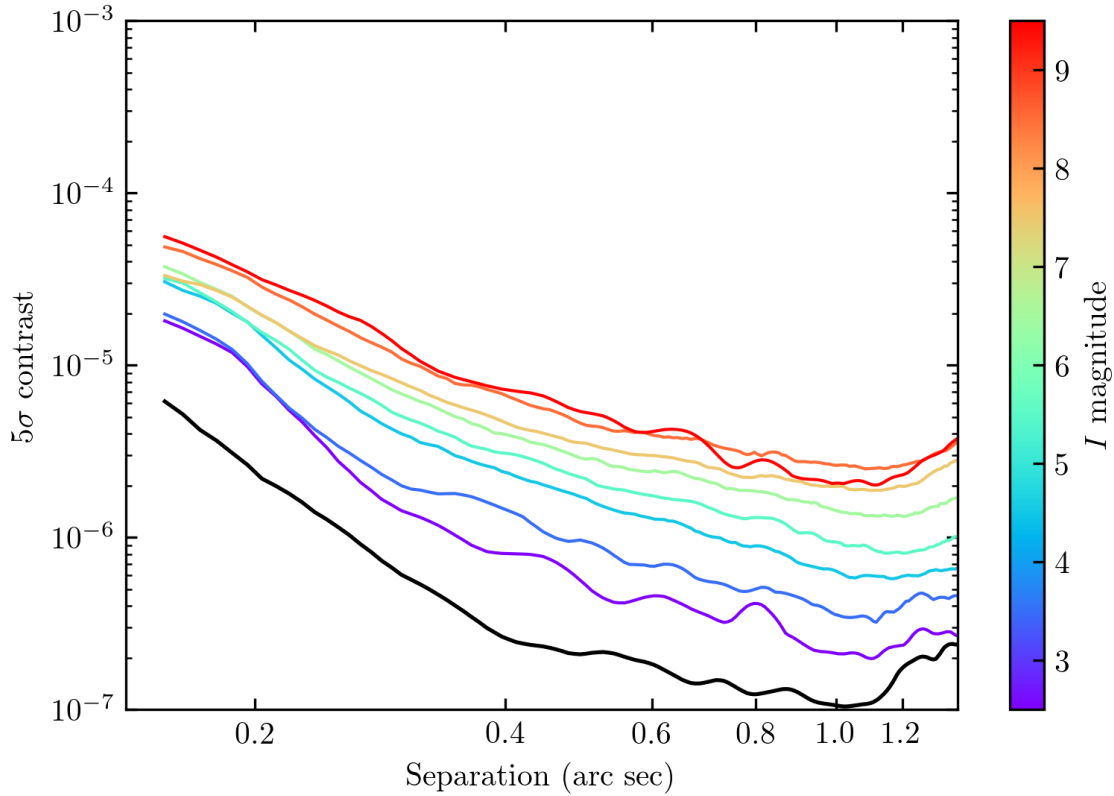

Figure 2: Median GPI detection thresholds from 40-minute *H*-band IFS PSF-subtracted observing sequences in the GPIES campaign. The lowermost black curve is for an 11-minute observation of Sirius.

The GPIES database[11] of campaign observations enables studies of correlations between performance and various instrument and environmental parameters. Previous studies have investigated the relationships between GPI AO residual wavefront error and contrast as well as seeing and contrast[19]. In this work we do not include information from AO telemetry; while this information adds predictive power, it is a metric for GPI performance, rather than an underlying cause. We have investigated correlations between post-processed H-band contrast and target parameters (stellar brightness, WFS flux), environment parameters (free-atmosphere $r_0$ and $\tau_0$, ambient windspeed and temperature, etc.) and/or observatory parameters (telescope pointing, telescope and instrument temperatures). We used multi-variate regression to fit final contrast as a function of $\tau_0$, seeing amplitude, ground windspeed, temperature difference between the GPI AO bench and ambient ($|\Delta T|$), star H-band magnitude, and airmass. To isolate the effects of these variables, only the subset of sequences taken in good observing conditions and with at least 1 Full Width Half Max-equivalent field rotation was used; for further description, see . This analysis and its implications for performance if GPI is moved to the Gemini North telescope are discussed in more detail in a companion paper[20].

As Figure 3 shows, the most predictive variables are $\tau_0$, $|\Delta T|$, star magnitude, and airmass; the addition of ground windspeed, seeing amplitude and total ADI field rotation adds negligible additional predictive power, as quantified by the $R^2$ statistic. From this, we can see that for a given stellar magnitude and airmass, GPI contrast is determined by free-atmosphere $\tau_0$ and (surprisingly) by dome/mirror seeing. The latter is often quite significant, as temperature differentials between the primary mirror and outside air can exceed 4 degrees Celsius; we examine this in more detail in a separate paper[3]. The negligible impact of ambient seeing amplitude is at first surprising; however, for a given spatial scale, the DMs can correct turbulence of any amplitude up to the stroke limit of the actuators. Hence, contrast at a fixed separation is uncorrelated with seeing amplitude until a threshold is crossed (~1.2"); GPI is not typically used above this threshold. The coherence time of the seeing, $\tau_0$, is the more important factor.

In addition, contrast continues to improve with good conditions and brighter targets rather than hitting a floor. This indicates that quasi-static speckle aberrations are well removed by KLIP processing, due to the low inherent level of non-commonpath aberrations, GPI's achromatic PSF, and relatively stable optical system. A similar analysis has been conducted using machine learning techniques[21].

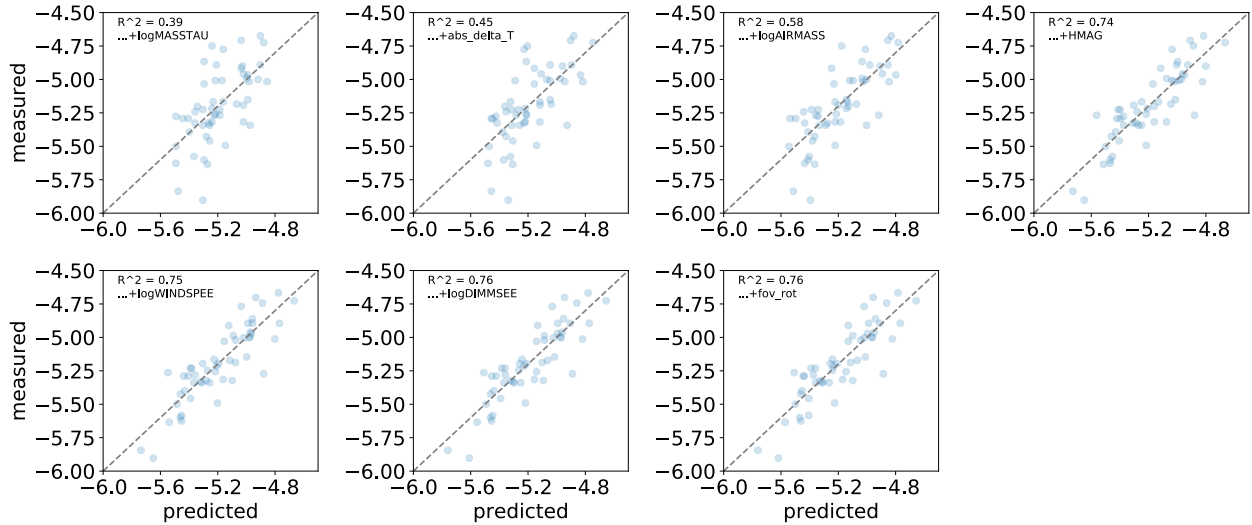

**Figure 3: Comparison of regression-based contrast predictions (X-axes) vs measured contrast (Y-axes) for the good-conditions subset of GPIES campaign sequences described above. Contrast is evaluated at 0.4" in complete post-processed GPIES observing sequences. (For this figure, these are based on a simple 5-sigma cutoff rather than the FMMF). Each successive panel adds a regression with a different variable. Top row: $\tau_0$, temperature differential between telescope/instrument and ambient air, airmass, H magnitude. Bottom row adds windspeed, DIMM-estimated seeing, and field of view rotation.**

The GPI PSF, like other high-contrast PSFs, includes an elongated or hourglass-shaped central component associated with servo-lag errors; we typically refer to this as the "butterfly" pattern. Analysis of the GPIES PSF database combined with NOAA Global Forecasting System data shows that the orientation of this butterfly feature correlates strongly with the direction and magnitude of the jet stream winds over Cerro Pachon[22], indicating that the jet stream is the likely cause of the relatively fast $\tau_0$ seen.

## 4. POTENTIAL UPGRADES

The GPIES campaign survey phase will conclude in 2018 with final follow-up of candidate companions in 2019. The Gemini Observatory is anticipating delivery of new instruments to Gemini South, prioritizing synergies with survey projects such as LSST. In that context, we are examining prospects for moving GPI to Maunakea. GPI's original design was fixed in 2008, and the field of AO and high-contrast imaging has progressed significantly since then with new technology such as faster computers, new approaches such as pyramid wavefront sensors, and new capabilities such as spectropolarimetry. We are developing plans for an upgrade dubbed GPI 2.0. The scientific motivation and potential architecture are discussed in more detail in [23] and briefly summarized here.

### 4.1 Scientific motivation

Upgrading and moving GPI only makes sense if a compelling scientific program can be identified, especially in the context of other instruments on Mauna Kea observatories. Merely repeating the GPIES survey in the north would result in perhaps 1-4 additional planetary or brown dwarf companions. However, with enhanced capabilities, GPI 2.0 can carry out a several new projects providing new insights into the properties and formation of giant exoplanets and circumstellar debris disks.

- It is possible that a significant number of planets (including Jupiter) formed through "cold-start" core accretion processes. Currently such planets are only detectable by GPI or SPHERE under the very best conditions and hence only for a small number of stars. Replicating best performance under typical conditions and generally enhancing bright-star contrast by a factor of ~4 would open up cold-start planet detection around hundreds of stars. Particularly when combined with GAIA astrometric observations this would allow statistical determination of the initial entropy of wide-orbit giant planets and hence constrain their formation mechanism.

- The greatest concentration of very young (<3 Myr) stars in the solar neighborhood is in the Taurus star forming region; possibly a site of active giant planet formation. However, stars in this association are typically too faint for current GPI, and the IWA of 0.15" corresponds to ~20 AU. Enhancing the limiting magnitude and IWA would allow the possibility of observations of newly-formed or actively-accreting planets

- An enhanced magnitude limit would allow large-scale surveys of asteroids and other faint solar system objects

- GPI photometry is currently accurate at the ~5% level. If this could be improved to ~1% in either absolute or relative spectrophotometric terms, we could search for variability in exoplanets due to rotation or time evolution of clouds

- Moderate (R~1000) to high (R~20,000) resolution spectroscopy of exoplanets can provide significant constraints on planetary atmospheric properties such as metallicity or gravity with less uncertainty due to modeling. High-resolution spectroscopy can also measure planet orbital velocities or rotation

- Giant exoplanets could potentially be polarized at the ~1% level at particular wavelengths. Precise calibration combined with a spectropolarimetric could identify this signature, constraining cloud properties or the oblateness of the planet.

### 4.2 Upgrade concepts

These science cases suggest a series of potential upgrades for GPI 2.0. The exact scope of the upgrade will depend on funding and timescales – since GPI is a cassegrain facility instrument it is difficult to do incremental upgrades and preferable to do a single refurbmishment in a off-site facility. Major new features could include

- A pyramid wavefront sensor, modeled on proposed designs for the TMT NFIRAOS NGS wavefront sensor, feeding an EMCCD detector. This would likely increase WFS sensitivity by ~3-4 magnitudes, allowing performance at I~13 mag. Pyramid sensors also provide some anti-aliasing protection and better non-common-path errors

- A new MEMS deformable mirror if one can be procured with no defective actuators in the clear aperture

- Replacement of the realtime AO computer with a higher-performance architecture, either conventional CPUs or a GPU-based approach. This could be combined with implementation of predictive control. The goal would be 2 kHz operation with ~0.5 ms total delay from start of CCD read.

- New APLC designs optimized for higher throughput and smaller IWA

- Replacement of the CAL interferometer with a fast focal-plane camera, for realtime speckle measurement and potentially focal-plane wavefront sensing (either directly or through a self-coherent-camera approach[23])

- A steerable fiber-feed module for sending light to a fixed high-resolution spectrograph

- An additional spectral prism providing broadband simultaneous YJHK operation, like the CHARIS instrument would be used for rapid surveys and higher SNR on faint planets. This could potentially be combined with a spectropolarimetric capability

- Operability enhancements so the instrument can be operated in traditional queue mode with full automation from acquisition to conclusion of observations

Several potential risks have been identified for the move to Gemini North. The most concerning is the telescope secondary mirror. The Gemini North M2 has significant print-through from support structures, six complex bumps of ~3 microns PV covering ~1 m patches. Simulations (Figure 4) show these producing scattered light features at the $10^{-4}$ level. In principal these would be stable and could be subtracted by ADI / SDI processing but likely would have higher-order effects. Simulations are ongoing to explore mitigation strategies, including precomputed phase offsets to the AO system and a custom Lyot mask.

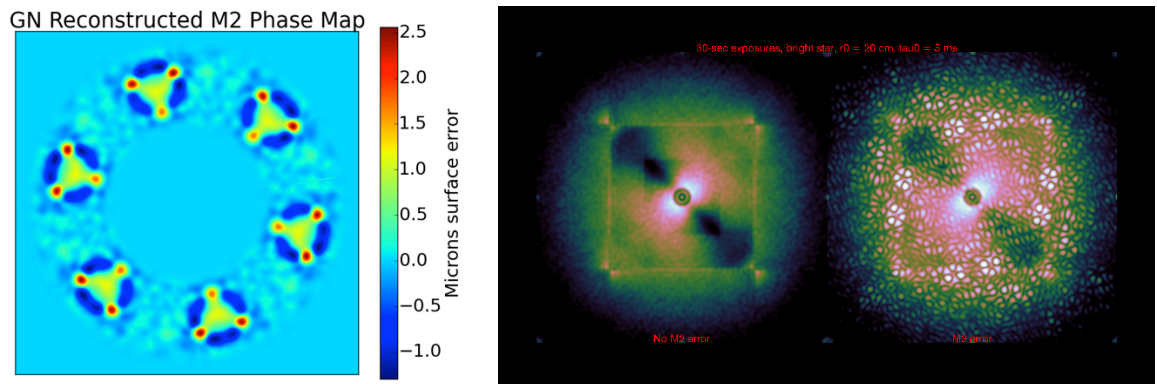

Figure 4: Left: Phase map of the Gemini North secondary mirror from phase diversity measurements (Veran & Lai, private communication). Right: simulated 30-second exposure PSFs for typical Mauna Kea seeing, with and without the M2 aberrations.

Overall, an upgraded GPI on Gemini North would have significant science capability and could remain productive until supplanted by ELT high-contrast imaging later in the next decade.